\documentclass[12pt]{article}
\usepackage{amsmath}
\usepackage{amsfonts}
\usepackage{bm}
\advance \textheight by \topskip
\advance \textheight by 1pt
\makeatother
\def\bea{\begin{eqnarray}}
\def\ena{\end{eqnarray}}
\def\non{\nonumber}

\newtheorem{prop}{Proposition}
\newtheorem{theorem}{Theorem}
\newtheorem{defn}{Definition}
\newtheorem{lemma}{Lemma}
\newtheorem{cor}{Corollary}

\newtheorem{remark}{Remark}

\newcommand{\bc}[2]{
\left(
\begin{array}{c}{#1}\\{#2}\end{array}
\right)}

\title{
Degeneration of Trigonal Curves and Solutions of the KP-Hierarchy
}

\author{
Atsushi Nakayashiki\thanks{
e-mail: atsushi@tsuda.ac.jp}\\
Department of Mathematics,
Tsuda University\\
Kodaira, Tokyo 187-8577, Japan
\\
\\
 \\
}

\date{}
\begin{document}

\maketitle
\thispagestyle{empty}
\vskip15mm

\begin{abstract}
It is known that soliton solutions of the KP-hierarchy corresponds to singular rational curves 
with only ordinary double points. 
In this paper we study the degeneration of theta function solutions corresponding to certain 
trigonal curves. We show that,  when the curves degenerate to singular rational curves with only ordinary triple points, the solutions tend to some intermediate solutions between solitons and rational solutions.
They are considered as cerain limits of solitons. 

The Sato Grassmannian is extensively used here to study the degeneration of solutions, since it directly connects solutions of the KP-hierarchy to the defining equations of algebraic curves.
We define a class of solutions in  the Wronskian form which contains soliton solutions as a subclass 
and prove that, using the Sato Grassmannian,  the degenerate trigonal solutions are connected to those solutions by certain gauge transformations
 
\end{abstract}
\clearpage
\pagestyle{plain}

\setcounter{page}{1}

\section{Introduction}
In this paper we study the limits of theta function solutions of the KP-hierarchy corresponding to
certain trigonal curves when they degenerate to singular rational curves with only ordinary triple points 
as their singularities. There are two major motivations in the present research.

Recently soliton solutions of the KP equation are extensively studied in connection with the various wave patterns of line solitons, total positivity of  the Grassmannians  and cluster algebras \cite{K,ChK1,ChK2,KW}. 
From the view point of spectral curves soliton solutions correspond to singular rational curves with 
 only ordinary double points \cite{AbG, TD, Man, Mum2, SW, Ab}.  Therefore it is interesting to study how the structure of soliton solutions 
 studied in \cite{K}  is reflected in that of theta function solutions corresponding to non-singular algebraic curves. To study this problem it is necessary to connect theta function solutions to soliton solutions 
 explicitly. In general  to compute explicitly the limits of theta function solutions is a non-trivial but curious problem as itself.
The case of hyperelliptic curves is relatively well studied \cite{BBEIM,Mum2}. In this case soliton solutions are 
obtained for example. However not all soliton solutions are obtained as limits of hyperelliptic solutions.  So it is important to study the case of non-hyperelliptic curves. But the problem is that 
to compute a limit of a non-hyperelliptic solution is not very easy.
The reason is that it is difficult to find a canonical homology basis explicitly for non-hyperelliptic curves in general.  
 
In the last two decades the theory of  multi-variate sigma functions associated with higher genus 
algebraic curves had been developped \cite{BEL1,BEL2,BEL3,N1,N2,N3,KS}. The higher genus sigma function 
is a certain modification of the Riemann's theta function.
The most important property of it is the modular invariance. It means that, although the sigma function is defined by specifying a canonical homology basis, it does not depend on the choice of it. 
For some class of algebraic curves, such as $(n,s)$ curves, the modular invariance 
is expressed in a more strong form. Namely the Taylor coefficients 
of the sigma function become polynomials of coefficients of the defining equations of an algebraic curve. 
It means that the sigma function has a limit when the coefficients of the defining equations of a curve are specialized in any way.
So it is an intriguing and a well-defined problem to establish the higher genus generalization of the ``elliptic-trigonometric-rational degeneration ''of Weierstarss elliptic function. The paper \cite{Ber-L} studies some  problem in this direction. 

In the present paper we study the degeneration of the theta function solutions corresponding to certain 
non-hyperelliptic curves. To this end we extensivley use the Sato Grassmannian \cite{SS}. It is the moduli space 
of the formal power series solutions of the KP-hierarchy (see Theorem 1).  Therefore a theta function solution
corresponds to some point of the Sato Grassmannian. The importance of the Sato Grassmannian here is in the fact  that it connects directly solutions of the KP-hierarchy to the defining equations of the algebraic curves. It means that one can compute the limit of a theta solution without information about canonical homology bases. 
This is the main idea in the present paper.

Now let us explain the content of the paper in more detail.
We consider the trigonal curve, which is  a special case of the so called a $(3,3m+1)$ curve \cite{BEL2,BL1},  given by the equation 
\bea
&&
y^3=x\prod_{j=1}^{3m}(x-\lambda_j^3),
\label{3s-curve}
\ena
 and its degeneration to the singular rational curve with only ordinary triple points:
\bea
&&
y^3=x\prod_{j=1}^m(x-\lambda_j^3)^3.
\label{3s-curve-degenerate}
\ena
A solution of the KP-hierarchy corresponding to (\ref{3s-curve}) can be expressed by the higher genus sigma function \cite{N2}. As mentioned above the sigma function has a well defined limit when the curve (\ref{3s-curve}) degenerates 
to the curve (\ref{3s-curve-degenerate}). Unfortunately we do not yet know the explicit form of the limit of the sigma function. The point is that we can compute the limit of the solution independently of the sigma function using Sato Grassmannian. Then, in turn, it will be possible to determine the limit of the sigma function using it.

The strategy to determine the limits of solutions is as follows.
We first define a class of solutions in the Wronskians form 
which contains soliton solutions as a subclass. We call 
a solution in this class a generalized soliton (GS). We embed GS's to the Sato Grassmannian. Next we embed
theta function solutions corresponding to the curves (\ref{3s-curve}) to the Sato Grassmannian.  As explained before  it is easy to take the degeneration limits of trigonal solutions in the Sato Grassmannian.
We then compare them 
with the GS's. We prove that a solution in the former is connected to a solution in the latter by a certain gauge transformation.
In this way the degenerate trigonal solutions are shown to be expressed by Wronskians of functions 
determined from the algebraic curves (\ref{3s-curve-degenerate}). They are some intermediate solutions between solitons and rational solutions which can be considered as certain degenerate 
limits of solitons.

The organization of the present paper is as follows.
In section 2 after a brief explanation on the KP-hierarchy,  
the Sato Grassmannian and the correspondence of points of it with solutions of the KP-hierarchy are reviewed. 
The Wronskian construction of $(n,k)$ solitons are reviewed in section 3.
 In section 4 generalized soliton 
solutions are introduced and the corresponding points in the Sato Grassmannian are determined.
An alternative description of GS's is given for the sake of applications in section 5.
 In section 6 an example of a frame of a GS which is relevant to the description of degenerate trigonal solutions is given. 
The map from the set of the affine rings of non-singular algebraic curves to 
the Sato Grassmannian is summarized in section 7. 
In section 8 the definition of the higher genus sigma function and the description of the solution corresponding 
to the affine ring of  the curve (\ref{3s-curve})  in terms of the sigma function are reviewed.
The degeneration of the 
sigma function solution is determined in section 9. In section 10 an example of the solution in the simplest case $m=1$ is given in detail.

\section{KP-hierarchy and the Sato Grassmannian}
The KP-hierarchy in the bilinear form \cite{DJKM} is the equation for the function $\tau(x)$, $x=(x_1,x_2,x_3,...)$ 
given by 
\bea
&&
\oint {\rm e}^{-2\sum_{j=1}^\infty y_j\lambda^j}\tau(x-y-[\lambda^{-1}])\tau(x+y+[\lambda^{-1}])\frac{d\lambda}{2\pi i}=0,
\label{KP-hierarchy}
\ena
where $y=(y_1,y_2,y_3,...)$, $[\lambda^{-1}]=(\lambda^{-1},\lambda^{-2}/2,\lambda^{-3}/3,...)$ and the integral siginifies 
taking the coefficient of $\lambda^{-1}$ in the series expansion. By expanding in $y$ it is equivalent to the 
infinite system of Hirota's bilinear equations. The first of which is the KP equation in the Hirota's 
bilinear form:
\bea
&&
(D_1^4-4D_1D_3+3D_2^2)\tau\cdot \tau=0.
\non
\ena
Here the Hirota derivatives $D_1^4$ etc. are defined by
\bea
&&
\tau(x+y)\tau(x-y)=\sum_{\alpha_i\geq 0} (D_1^{\alpha_1}D_2^{\alpha_2}\cdots)(\tau\cdot\tau)
\frac{y_1^{\alpha_1}y_2^{\alpha_2}\cdots}{\alpha_1!\alpha_2!\cdots}.
\non
\ena

If we set $u=2\partial_{x}^2\log \tau(x)$, $(x=x_1, y=x_2, t=x_3)$ we have the KP equation 
\bea
&&
3u_{yy}+(-4u_t+6uu_x+u_{xxx})_x=0.
\non
\ena
In this paper we mean by the KP hierarchy the equation (\ref{KP-hierarchy}) for $\tau(x)$.

Next we recall the definition of the Sato Grassmannian \cite{SS} which we denote by  UGM (universal Grassmann manifold) \cite{S,SN}(see also \cite{Mul}\cite{KNTY}) and the correspondence 
between solutions of the KP-hierarchy and points of UGM. 

Let $V={\mathbb C}((z))$ be the vector space of formal Laurent series in the variable $z$ and 
$V_\phi={\mathbb C}[z^{-1}]$, $V_0=z{\mathbb C}[[z]]$ subspaces of $V$. Then 
\bea
&&
V=V_\phi \oplus V_0,
\hskip5mm
V/V_0\simeq V_\phi.
\non
\ena
Let $\pi: V\rightarrow V_\phi$ be the projection map. The Sato Grassmannian UGM
is defined by
\bea
&&
{\rm UGM}=\{ \text{ a subspace $U$ of $V$}| \dim {\rm Ker}(\pi|_U)=\dim {\rm Coker}(\pi|_U)<\infty \}.
\non
\ena

An ordered basis of a point $U$ of UGM is called a frame of $U$.  We can express a frame of $U$ by an infinite matrix as follows.

We set 
\bea
&&
f_i=z^{i+1},\hskip5mm i\in {\mathbb Z},
\non
\ena
and write an element $f$ of $V$ as
\bea
&&
f=\sum_{i\in {\mathbb Z}} X_i f_i.
\label{f}
\ena

We associate the infinite column vector $(X_i)_{i\in{\mathbb Z}}$ to $f$.
Then a matrix $X=(X_{i,j})_{i\in {\mathbb Z},j\in {\mathbb N}}$ is a frame of $U$ if 
$(X_{i,j})_{i\in {\mathbb Z}}$, $j\in {\mathbb N}$ ia a basis of $U$. 
We write it as 
\bea
&&
X=
\left(
\begin{array}{ccc}
\quad&\vdots&\vdots\\
\cdots&X_{-2,2}&X_{-2,1}\\
\cdots&X_{-1,2}&X_{-1,1}\\
---&---&---\\
\cdots&X_{0,2}&X_{0,1}\\
\cdots&X_{1,2}&X_{1,1}\\
\quad&\vdots&\vdots\\
\end{array}
\right),
\ena
where the columns are labeled from right to left as $1,2,3,...$ and rows are labeled from 
up to down as $...,-1,0,1,...$.

For a point $U$ of UGM there exists
a frame  $X=(X_{i,j})_{i\in {\mathbb Z},j\in {\mathbb N}}$ of $U$ satisfying the following conditions:
there exists a non-negative integer $l$ such that
\bea
&&
X_{i,j}=\left\{
\begin{array}{cl}
1&\text{ if $j>l$ and $i=-j$ }\\
0&\text{ if ($j>l$ and $i<-j$) or ($j\leq l$ and $i<-l$)}.
\end{array}
\right.
\label{frame-cond}
\ena

It means that $X$ is of the form
\bea
&&
X=\left[\begin{array}{cccc}
\ddots&\quad&O&\quad\\
\cdots&1&\quad&\quad\\
\cdots&\ast&1&\quad\\
\cdots&\ast&\ast&B\\
\end{array}
\right],
\non
\ena
where $B$ is an $\infty\times l$ matrix and its first row is placed at the $-l$th row of $X$.
In the following  a frame of a point of UGM is always assumed to satisfy the condition (\ref{frame-cond}).

Here we introduce the notion of Maya diagram. 
A Maya diagram of charge $p$ is a sequence of integers $M=(m_1,m_2,...)$  such that
$m_1>m_2>\cdots$ and, for some $l$, $m_i=-i+p$, $i\geq l$.
In this paper we consider only a Maya diagram of charge $0$ and call them simply a Maya diagram. 
With each Maya diagram $M$ is associated the partition $\lambda$ by
\bea
&&
\lambda=(m_1+1,m_2+2,...),
\non
\ena
and vice versa.

Let $\lambda$ be an arbitrary partition and $M=(m_1,m_2,m_3,...)$ the Maya diagram corresponding to it. 
The Pl\"ucker coordinate $X_\lambda$ of a frame $X$ is defined as 
\bea
&&
X_\lambda=\det(X_{m_i,j})_{i,j\in {\mathbb N}},
\label{det-1}
\ena
which is also denoted by $X_M$ using the Maya diagram $M$.

The infinite determinant (\ref{det-1}) is actually defined as a finite determinant as follows.
Let $l$ be an integer in the condition (\ref{frame-cond}) and $l_1$ an integer such that $m_i=-i$, $i\geq l_1$. Take an integer $l_2$ such that $l_2\geq \max(l+1,l_1)$.
Then 
\bea
&&
X_\lambda=\det(X_{m_i,j})_{1\leq i,j\leq l_2}.
\label{det-2}
\ena
It does not depend on the choice of $l, l_1, l_2$.

For a frame $X$ satisfying  (\ref{frame-cond}) we define the $\tau$ function by
\bea
&&
\tau(x;X)=\sum_\lambda X_\lambda s_\lambda(x),
\label{schur-exp}
\ena
where the summation is taken over all partitions and $s_\lambda(x)$ is the Schur function 
corresponding to the partition $\lambda$ (see the next section).

A frame $X$ satisfying the condition (\ref{frame-cond}) is not unique for $U$. If $X$ is replaced with 
another frame  $\tau(x;X)$ is multiplied by a non-zero constant.

\begin{theorem}\label{Sato1}{\rm \cite{SS}} For a frame $X$ of a point $U$ of UGM,  $\tau(x,X)$ is a solution of the KP-hierarchy. Conversely for any formal 
power series solution $\tau(x)$ of the
KP-hierarchy there exists a unique point $U$ of ${\rm UGM}$ such that 
$\tau(x)=c(X)\tau(x;X)$ for a frame $X$ of $U$ and some constant $c(X)$ which depends on $X$.
\end{theorem}

\section{$(n,k)$-soliton}
Let us recall soliton solutions of the KP-hierarchy.

We first recall the general Wronskian construction of solutions of the KP-hierarchy\cite{FN,S}.
Let $f_1,...,f_m$ be functions which satisfy 
\bea
&&
\frac{\partial f_l}{\partial x_j}=\frac{\partial^j f_l}{\partial x_1^j},
\hskip5mm
l\geq 1.
\label{LDE}
\ena
Then 
\bea
&&
\tau(x)={\rm Wr}(f_1,...,f_m)=\det(f^{(i-1)}_j)_{1\leq i,j\leq m},
\hskip5mm
f^{(i)}_j=\frac{\partial^i f_j}{\partial x_1^i},
\non
\ena
is a solution of the KP-hierarchy.

Using Wronskian method the soliton solutions can be constructed in the following way.

Let $n$ and $k$ be positive integers such that $n>k$, $\lambda_1,...,\lambda_n$ mutually 
distinct non-zero complex numbers and $A=(a_{ij})_{1\leq i\leq n,1\leq j\leq k}$ an $n\times k$ 
matrix of rank $k$. We set
\bea
&&
f_j=\sum_{i=1}^n a_{ij} e^{\xi_i},
\hskip5mm
\xi_i=\xi(x,\lambda_i),
\qquad
\xi(x,\lambda)=\sum_{j=1}^\infty x_j \lambda^j.
\label{S-function}
\ena
Since $e^{\xi_i}$ satisfies (\ref{LDE}) so does $f_j$. Therefore 
\bea
&&
\tau(x)={\rm Wr}(f_1,...,f_k)=\sum_{I=(i_1,...,i_k), 1\leq i_1<\cdots<i_k\leq n} \Delta_I(\lambda) A_I e^{\xi_{i_1}+\cdots+\xi_{i_k}},
\label{sato-soliton}
\ena
is a solution of the KP-hierarchy, where, for $I=(i_1,...,i_k)$,
\bea
&&
\Delta_I(\lambda)=\prod_{p<q}(\lambda_{i_q}-\lambda_{i_p}),
\quad
A_I=\det(a_{i_p,q})_{1\leq p, q\leq k}.
\non
\ena

\begin{defn}
We call the solution (\ref{sato-soliton}) an $(n,k)$ soliton.
\end{defn}

Sato determined the point of UGM corresponding to  an $(n,k)$ soliton.

\begin{theorem}\label{S-frame} {\rm \cite{S}}
The point of UGM corresponding to (\ref{sato-soliton}) is given by the frame which consists 
of the following functions:
\bea
&&
z^{-(k-1)}\sum_{i=1}^n \frac{a_{ij}}{1-\lambda_i z} \quad (1\leq j\leq k),
\hskip10mm
z^{-l} \quad (l\geq k),
\label{soliton-frame}
\ena
where $1/(1-\lambda_i z)$ is considered as a power series in $z$ by 
\bea
&&
\frac{1}{1-\lambda_i z}=\sum_{r=0}^\infty \lambda_i^r z^r.
\non
\ena
\end{theorem}

\section{Generalization of $(n,k)$-soliton}
In this section we define a class of solutions of the KP-hierarchy in Wronskian form which contains 
$(n,k)$ solitons.

Let $n, k$ as before, $N$,  $r_j$ $(0\leq j\leq N)$ non negative integers such that 
\bea
&&
r_0+\cdots+r_N=n,
\non
\ena
 $\lambda_{i,j}$, $0\leq i\leq N$, $1\leq j\leq r_i$ non-zero complex numbers such that, for each
$i$, $\lambda_{i,j}\neq \lambda_{i,j'}$ if $j\neq j'$, and $A=(a_{ij})$ an $n\times k$ matrix 
of rank $k$. We set $r_{-1}=0$ for convenience. We aslo use $\lambda_i$, $1\leq i\leq n$,  defined by
\bea
&&
(\lambda_1,...,\lambda_n)=(\lambda_{0,1},...,\lambda_{0,r_0},...,\lambda_{N,1},...,\lambda_{N,r_N}).
\non
\ena
Set 
\bea
&&
f_j=\sum_{s=0}^N\sum_{i=r_{-1}+\cdots+r_{s-1}+1}^{r_0+\cdots+r_s} a_{ij}
\left(\frac{d^s}{d \lambda^s} e^{\xi(x,\lambda)}\right)_{\lambda=\lambda_i}.
\quad
1\leq j\leq k.
\label{GS-function}
\ena
Since the derivatives in $x_i$ and $\lambda$ commute, $f_j$ still satisfies Equation (\ref{LDE}).
Therefore the Wronskian of the functions $f_j$ becomes a 
solution of the KP-hierarchy. 

\begin{defn}
The solution $\tau(x)={\rm Wr}(f_1,...,f_k)$ with $f_j$ given by  (\ref{GS-function}) is 
called an $(n,k)$ generalized soliton(GS).
\end{defn}

\begin{remark} The above construction of $(n,k)$ GS's is similar to that of rational solutions of 
KP equation in \cite{Kr1,Kr2,Wil}.
\end{remark}

The point of UGM corresponding to an $(n,k)$ GS can be determined as follows.
Let
\bea
&&
{\mathbf v}^{(i)}(\lambda,z)=\frac{d^i}{d \lambda^i}\frac{1}{1-\lambda z}=\frac{i!z^i}{(1-\lambda z)^{i+1}},
\quad
i\geq 0.
\label{v-i}
\ena

\begin{theorem}\label{frame-GS} The point of UGM corresponding to an $(n,k)$ GS specified by 
(\ref{GS-function}) is given by the following frame:
\bea
&&
z^{-(k-1)}\left(
\sum_{s=0}^N\sum_{i=r_{-1}+\cdots+r_{s-1}+1}^{r_0+\cdots+r_s} a_{ij} {\mathbf v}^{(s)}(\lambda_i,z)
\right), 
\quad 1\leq j\leq k,
\non
\\
&&
z^{-j}, \quad  j\geq k.
\label{GSF}
\ena
\end{theorem}

\noindent
{\it Proof.} The proof is similar to Sato's proof of Theorem \ref{S-frame}.

The linear independence of the functions (\ref{GSF}) is equivalent to ${\rm rank } \,A=k$. 
Therefore they define the frame of a point of UGM. We shall show that the solution 
 (\ref{schur-exp}) corresponding to the frame  (\ref{GSF}) is equal to the 
 Wronskian of the set of  functions $f_1,...,f_k$. To this end we first express the frame 
  (\ref{GSF}) in the matrix form and compute the Plucker coordinates of it.

Let $X=(X_{i,j})_{i\in {\mathbb Z}, j\in {\mathbb N}}$ be the infinite matrix representing
the frame (\ref{GSF}). It is described in the following way.

We define the vectors ${\mathbf u}^{(i)}(\lambda)$, $i\geq 0$ with the infinite components by
\bea
&&
{\mathbf u}^{(i)}(\lambda)
=
\frac{d^i}{d \lambda^i}
\left[\begin{array}{c}1\\ \lambda \\ \lambda^2\\ \vdots\end{array}\right].
\non
\ena
The components of  ${\mathbf u}^{(i)}(\lambda)$ are labeled from up to down as $0,1,2,...$. 
Let $C$ and $B$ be the ${\mathbb Z}_{\geq 0}\times n$ and ${\mathbb Z}_{\geq 0}\times k$ 
matrices defined respectively by
\bea
C&=&\left({\mathbf u}^{(0)}(\lambda_{01}),...,{\mathbf u}^{(0)}(\lambda_{0,r_0}),...,
{\mathbf u}^{(N)}(\lambda_{N,1}),...,{\mathbf u}^{(N)}(\lambda_{N,r_N})\right),
\non
\\
B&=&C A=(b_{i,j})_{i\in {\mathbb Z}_{\geq 0}, 1\leq j\leq k}.
\non
\ena
Then $X$ is given by
\bea
&&
X_{i,j}=\left\{
\begin{array}{ll}
\delta_{i,-j}& \text{if $j\geq k+1$}\\
b_{i+k, k+1-j}& \text{if $1\leq j\leq k$ and $i\geq -k$}\\
0&\text{ if  $1\leq j\leq k$ and $i<-k$}
\end{array}
\right.
\non
\ena
Namely $X$ is of the form
\bea
&&
X=\left[\begin{array}{cccc}
\ddots&\quad&\quad&\quad\\
\quad&1&\quad&\quad\\
\quad&\quad&1&\quad\\
\quad&\quad&\quad&B\\
\end{array}
\right],
\non
\ena
where the $0$th row of $B$ sits at the $-k$th row of $X$.

We consider the Plucker coordinate $X_M$ of $X$ corresponding to a Maya diagaram $M$.
The form of $X$ given above implies that $X_M=0$ unless $M$ is written in the form
\bea
&&
M=(m_1,...,m_k, -k-1,-k-2,...), 
\quad
m_1>\cdots>m_k\geq -k.
\label{X-Maya}
\ena
Let 
\bea
&&
{\cal L}=\{(l_l,...,l_k) | 0\leq l_1<\cdots<l_k\}.
\non
\ena
We associate the element $(l_1,...,l_k)$ of ${\cal L}$ with $M$ by
\bea
&&
(l_1,...,l_k)=(m_k+k,...,m_1+k).
\non
\ena
By this map the set of Maya diagrams of the form (\ref{X-Maya}) and ${\cal L}$ bijectively correspond.
For $(l_l,...,l_k)\in {\cal L}$ we set $B_{l_1,...,l_k}:=\det(b_{l_i,j})_{1\leq i,j\leq k}$.
If  $(l_l,...,l_k)\in {\cal L}$ corresponds to $M$ then it is easy to see that
\bea
&&
X_M=B_{l_1,...,l_k}.
\non
\ena
Let $\mu$ be the partition corresponding to $M$:
\bea
&&
\mu=(m_1+1,...,m_k+k),
\non
\ena
which is related with $(l_l,...,l_k)$ by
\bea
&&
\mu=(l_k-(k-1),...,l_2-1,l_1).
\label{mu-l-corresp}
\ena
Let $p_j(x)$ be the polynomial of $x=(x_1,x_2,...)$ defined by
\bea
&&
{\rm e}^{\xi(x,\lambda)}=\sum_{j=0}^\infty p_j(x)\lambda^j.
\non
\ena
We set $p_j(x)=0$ for $j<0$.
In terms of $(l_1,...,l_k)$ the Schur function $s_\mu(x)$ is given by
\bea
&&
s_\mu(x)=\det(p_{l_j-i}(x))_{0\leq i\leq k-1, 1\leq j\leq k}.
\label{Schur-def}
\ena
We sometimes denote $s_\mu(x)$ by $s_{l_1,...,l_k}(x)$.
For $(l_1,...,l_k)\in {\cal L}$ and $1\leq r_1<\cdots<r_k\leq n$ we set
\bea
C_{l_1,...,l_k}^{r_1,...,r_k}&=&\det(C_{l_i,r_j})_{1\leq i,j\leq k},
\non
\\
A_{r_1,...,r_k}&=&\det(a_{r_i,j})_{1\leq i,j\leq k}.
\non
\ena
Then we have, by (\ref{schur-exp}) and the Cauchy-Binet formula for the determinant, 
\bea
\tau(x)&=&
\sum_{0\leq l_1<\cdots<l_k} B_{l_1,...,l_k}s_{l_1,...,l_k}(x)
\label{CB-1}
\\
&=&
\sum_{0\leq l_1<\cdots<l_k}
\sum_{1\leq r_1<\cdots<r_k\leq n}
C_{l_1,...,l_k}^{r_1,...,r_k}A_{r_1,...,r_k}s_{l_1,...,l_k}(x).
\non
\ena
Let us rewrite the last expression. Let $P=(p_{j-i}(x))_{0\leq i\leq k-1, j\in {\mathbb Z}_{\geq 0}}$.  
 We denote the $r$th column vector of $C$ by $C_r$.

\begin{lemma}\label{exchange-sum}
The following equation is valid:
\bea
&&
\sum_{0\leq l_1<\cdots<l_k}
C_{l_1,...,l_k}^{r_1,...,r_k}s_{l_1,...,l_k}(x)
=\det\left(P\cdot (C_{r_1},...,C_{r_k})\right).
\label{Cauchy-Binet}
\ena
\end{lemma}
\vskip2mm
\noindent
{\it Proof.} By the Cauchy-Binet formula and (\ref{Schur-def}) the right hand side of  (\ref{Cauchy-Binet})
is written as
\bea
&&
\sum_{0\leq l_1<\cdots<l_k}
\det\left(p_{l_j-i}(x)\right)_{0\leq i\leq k-1,1\leq j \leq k}C_{l_1,...,l_k}^{r_1,...,r_k}
=
\sum_{0\leq l_1<\cdots<l_k}
s_{l_1,...,l_k}(x) C_{l_1,...,l_k}^{r_1,...,r_k}
\non
\ena
which shows (\ref{Cauchy-Binet}). $\Box$

Let us compute the produc of matrices inside the determinant in the right hand side of 
(\ref{Cauchy-Binet}).

Since 
\bea
&&
\left(P{\mathbf u}^{(0)}(\lambda)\right)_s=\sum_{j=0}^\infty p_{j-s}(x)\lambda^j
=\lambda^s\sum_{j=s}^\infty p_{j-s}(x)\lambda^{j-s}=\lambda^s{\rm e}^{ \xi(x,\lambda)},
\non
\ena
we have 
\bea
&&
\left(P{\mathbf u}^{(i)}(\lambda)\right)_s=\left(P \frac{d^i}{d\lambda^i}{\mathbf u}^{(0)}(\lambda)\right)_s
=\left(\frac{d^i}{d\lambda^i} P {\mathbf u}^{(0)}(\lambda)\right)_s
=\frac{d^i}{d\lambda^i}\left(\lambda^s {\rm e}^{\xi(x,\lambda)}\right).
\non
\ena

Let 
\bea
&&
\tilde{{\mathbf u}}^{(i)}(x,\lambda)=
\frac{d^i}{d \lambda^i} \left(\left[\begin{array}{c}1 \\ \lambda
 \\ \vdots\\ \lambda^{k-1}\end{array}\right] {\rm e}^{\xi(x,\lambda)} \right),
 \non
 \\
 &&
 {\tilde C}(x)=\left(
 \tilde{{\mathbf u}}^{(0)}(x,\lambda_{01}),..., \tilde{{\mathbf u}}^{(0)}(x,\lambda_{0r_0}),...,
 \tilde{{\mathbf u}}^{(N)}(x,\lambda_{N1}),..., \tilde{{\mathbf u}}^{(N)}(x,\lambda_{N,r_N})
 \right).
 \non
 \ena
 
 We denote the $r$th column vector of  ${\tilde C}(x)$ by ${\tilde C}(x)_r$. Then
 \bea
 &&
 P\cdot\left(C_{r_1},...,C_{r_k}\right)=\left( {\tilde C}(x)_{r_1},..., {\tilde C}(x)_{r_k}\right).
 \non
 \ena
 Exchanging the order of the summation in (\ref{CB-1}) , using Lemma \ref{exchange-sum} and 
 the Cauchy-Binet formula again we get
 \bea
 \tau(x)&=&\sum_{1\leq r_1<\cdots<r_k\leq n}A_{r_1,...,r_k}
 \det\left( {\tilde C}(x)_{r_1},..., {\tilde C}(x)_{r_k}\right)
 \non
 \\
 &=&
 \det\left( {\tilde C}(x)A\right).
 \non
 \ena
 
 The $j$th column vector of  ${\tilde C}(x)A$ is given by
 
 \bea
 \left( {\tilde C}(x)A\right)_j &=&
 \sum_{s=0}^N\sum_{i=r_{-1}+\cdots+r_{s-1}+1}^{r_0+\cdots+r_s} 
 a_{ij}  \tilde{{\mathbf u}}^{(s)}(x,\lambda_{i})
\non
\\
&=&
 \sum_{s=0}^N\sum_{i=r_{-1}+\cdots+r_{s-1}+1}^{r_0+\cdots+r_s} 
 a_{ij}
 \left(
 \frac{d^s}{d \lambda^s} \left[\begin{array}{c}{\rm e}^{\xi(x,\lambda)} \\ \lambda e^{\xi(x,\lambda)}
 \\  \vdots\\ \lambda^{k-1} e^{\xi(x,\lambda)}\\\end{array}\right]  
 \right)_{\lambda=\lambda_i}
 \non
\\
&=&
 \sum_{s=0}^N\sum_{i=r_{-1}+\cdots+r_{s-1}+1}^{r_0+\cdots+r_s} 
 a_{ij}
 \left(
 \frac{d^s}{d \lambda^s} \left[\begin{array}{c}{\rm e}^{\xi(x,\lambda)} \\ \frac{d}{dx} e^{\xi(x,\lambda)}
 \\  \vdots \\ \frac{d^{k-1}}{dx^{k-1}} e^{\xi(x,\lambda)} \end{array}\right]  
 \right)_{\lambda=\lambda_i}
  \non
\ena

\bea
&=&
 \sum_{s=0}^N\sum_{i=r_{-1}+\cdots+r_{s-1}+1}^{r_0+\cdots+r_s} 
 a_{ij}
 \left[\begin{array}{c}  \left(\frac{d^s}{d \lambda^s} {\rm e}^{\xi(x,\lambda)}\right)_{\lambda=\lambda_i}
 \\
  \frac{d}{dx}  \left( \frac{d^s}{d \lambda^s}e^{\xi(x,\lambda)} \right)_{\lambda=\lambda_i}
 \\ 
  \vdots 
 \\
   \frac{d^{k-1}}{dx^{k-1}} \left(\frac{d^s}{d \lambda^s} e^{\xi(x,\lambda)} \right)_{\lambda=\lambda_i}\end{array}\right]  
 \non
 \\
 &=&
 \left[\begin{array}{c} f_j \\ f_j'
 \\ \vdots\\ f_j^{(k-1)}\end{array}\right].
 \non
 \ena
 
 Thus 
 \bea
 &&
 \tau(x)=\det\left( {\tilde C}(x)A\right)=
 {\rm Wr}(f_1,...,f_k).
 \non
 \ena
 $\Box$

\section{Change of a basis}
For applications it is necessary to write $1/(1-\lambda z)^r$ as a linear combination of
$\{{\mathbf v}^{(i)}(\lambda,z)\}$. 

\begin{lemma} For $m\geq 0$ 
\bea
&&
\frac{1}{(1-\lambda z)^m}=\sum_{i=0}^m\frac{1}{i!}\bc{m}{i}\lambda^i {\mathbf v}^{(i)}(\lambda,z).
\label{base-change}
\ena

\end{lemma}
\vskip2mm
\noindent
{\it Proof.} By differentiate the equation
\bea
&&
\frac{1}{1-w}=\sum_{n=0}^\infty w^n
\non
\ena
$m$ times we get
\bea
&&
\frac{m!}{(1-w)^{m+1}}=\sum_{n=0}^\infty (n+m)(n+m-1)\cdots(n+1) w^n.
\non
\ena
Sunstituting $w=\lambda z$ and dividing by $m!$ we obtain
\bea
\frac{1}{(1-\lambda z)^{m+1}}
&=&\frac{1}{m!}\sum_{n=0}^\infty (n+m)\cdots(n+1) \lambda^n z^n
\non
\\
&=&\frac{1}{m!} \frac{d^m}{d \lambda^m} \sum_{n=0}^\infty \lambda^{n+m} z^n
\non
\\
&=& \frac{1}{m!} \frac{d^m}{d \lambda^m}\left(\lambda^m \frac{1}{1-\lambda z}\right).
\non
\ena
By computing the last expression using the Leibniz rule we get the desired result. $\Box$

The case of $m=0,1$ of (\ref{base-change}) is
\bea
\frac{1}{1-\lambda z}&=&{\mathbf v}^{(0)}(\lambda,z),
\non
\\
\frac{1}{(1-\lambda z)^2}&=&{\mathbf v}^{(0)}(\lambda,z)+\lambda {\mathbf v}^{(1)}(\lambda,z).
\label{base-change-example}
\ena

\begin{cor}\label{rational-func-GS} The set of frames of $(n,k)$ generalized solitons, $n\geq 1$, coincide with the 
set of frames of the following form:
\bea
&&
z^{-(k-1)}\frac{g_i(z)}{f_i(z)}, \hskip5mm 1\leq i\leq k,
\non
\\
&&
z^{-i}, \hskip5mm i\geq k,
\non
\ena
where $f_i(z)$, $g_i(z)$ are polynomials such that $f_i(0)\neq 0$, $\deg f_i(z)>\deg g_i(z)$ for any $i$
and  $\{g_i(z)/f_i(z)|1\leq i\leq k\}$ is linearly independent.
\end{cor}

The number $n$ in the corollary is determined from the structure of roots of $f_i(z)$ as one sees
in the example in the next section.

\section{Example of GS}
Let $a_{ij}$, $b_{ij}$, $1\leq i\leq n$, $1\leq j\leq k$ be complex numbers.

\begin{lemma}\label{2n-k-GS}
The following set of functions is a frame of a $(2n,k)$ GS if the first $k$ functions are linearly independent.
\bea
&&
z^{-k+1}\sum_{i=1}^n\left( \frac{a_{ij}}{1-\lambda_i z}
+ \frac{b_{ij}}{(1-\lambda_i z)^2}\right), \hskip5mm 1\leq j \leq k
\non
\\
&&
z^{-i}, \hskip5mm i\geq k.
\label{GSF-2n-k}
\ena
The corresponding matrix $A$ is given by 
\bea
&&
A=\left(
\begin{array}{ccc}
a_{11}+b_{11}&\ldots&a_{1k}+b_{1k}\\
\vdots&\vdots&\vdots\\
a_{n1}+b_{n1}&\ldots&a_{nk}+b_{nk}\\
\lambda_1b_{11}&\ldots&\lambda_1b_{1k}\\
\vdots&\vdots&\vdots\\
\lambda_nb_{n1}&\ldots&\lambda_nb_{nk}\\
\end{array}
\right).
\label{A-example}
\ena
\end{lemma}
\vskip2mm
\noindent
{\it Proof.} By Corollary \ref{rational-func-GS} the set of functions (\ref{GSF-2n-k}) is a frame 
of a GS under the assumtion of the lemma. Let us calculate the type of the GS and the matrix $A$ associated with it.

By (\ref{base-change-example}) we get 
\bea
&&
\sum_{i=1}^n \left(\frac{a_i}{1-\lambda_iz}+\frac{b_i}{(1-\lambda_i z)^2}\right)
\non
\\
&=&
\left[{\mathbf v}^{(0)}(\lambda_1,z),\ldots,{\mathbf v}^{(0)}(\lambda_n,z), {\mathbf v}^{(1)}(\lambda_1,z),\ldots,
{\mathbf v}^{(1)}(\lambda_n,z)\right]
\left[\begin{array}{c}a_1+b_1\\ \vdots\\ a_n+b_n\\ \lambda_1b_1\\ \vdots\\ \lambda_nb_n\end{array}\right].
\non
\ena
Therefore, by Theorem \ref{frame-GS},  (\ref{GSF-2n-k}) is a frame of a $(2n,k)$ GS with 
$N=1$, $r_0=r_1=n$, $\lambda_{0i}=\lambda_i$, $\lambda_{1i}=\lambda_i$ and the
$2n\times k$ matrix $A$ is given by (\ref{A-example}).
$\Box$

The solution $\tau(x)$ corresponding to the frame  (\ref{GSF-2n-k}) is written 
in the Wronskian form by the definition of a GS. Let us compute it.

\vskip2mm
Let ${\mathbf v}_k$, ${\mathbf w}_k$ be the column vectors with $k$ components given by 
\bea
{\mathbf v}_k(\lambda)&=&(\lambda^{i-1})_{1\leq i\leq k}
\non
\\
{\mathbf w}_k(\lambda)&=&(i\lambda^{i-1}+\lambda^i\xi'(x,\lambda))_{1\leq i\leq k},
\hskip5mm
\xi'(x,\lambda)=\frac{d \xi(x,\lambda)}{d \lambda}=\sum_{j=1}^\infty j \lambda^{j-1}x_j.
\non
\ena

Define the $k\times 2n$ matrix $D$ by
\bea
&&
D=({\mathbf v}_k(\lambda_1),\ldots,{\mathbf v}_k(\lambda_n),{\mathbf w}_k(\lambda_1),\ldots,
{\mathbf w}_k(\lambda_n))
\non
\ena

Then
\bea
&&
\tau(x)=\sum_{1\leq i_1<\cdots<i_k\leq 2n}A_{i_1,...,i_k}D_{i_1,...,i_k}e^{\xi_{i_1}+\cdots+\xi_{i_k}}
\label{2n-k-GSF-tau}
\ena
where $D_{i_1,...,i_k}$ is the minor determinants corresponding to 
columns $i_1,...,i_k$ of $D$. Notice that $\lambda_{n+j}=\lambda_j$ and  $\xi_{n+j}=\xi_j$ for
$1\leq j\leq n$. 
It is different from a soliton solution, since
$D_{i_1,...,i_k}$ depends on $x_i$'s in general.

\section{Affine rings and the Sato Grassmannian}
In this section we review how to associate a point of UGM with the affine ring of a compact Riemann surface.

Let $X$ be a compact Riemann surface of genus $g$, $p_\infty$ a point of $X$ and $z$ a local coordinate at $p_\infty$. We denote by $U_a$ the vector space 
of meromorphic functions on $X$ which are holomorphic on $X\backslash\{p_\infty\}$.
 It is called the affine ring of $X\backslash\{p_\infty\}$.

We define the map
\bea
&&
\iota:U_a\longrightarrow {\mathbb C}((z))
\non
\ena
 as follows.
 
 Let $F$ be an element of $U_a$ and $F(z)$ the Laurent expansion of $F$ at $p_\infty$ in $z$.
Then we define
\bea
&&
\iota(F)=z^g F(z).
\non
\ena

\begin{theorem}{\rm \cite{SW,Mul,N2}}
The subspace $\iota(U_a)$ belongs to UGM.
\end{theorem}

\section{$(3,3m+1)$ Curves and the sigma function solution}
Let $m$ be a positive integer  and  $\lambda_1$,...,$\lambda_{3m}$ non-zero complex numbers such that  
$\lambda_1^3$,...,$\lambda_{3m}^3$ are mutually distinct. 
 We consider the algebraic curve defined by 
\bea
&&
y^3=x\prod_{j=1}^{3m}(x-\lambda_j^3).
\label{nonsing-3s}
\ena
It can be compactified by adding one point at $\infty$. We denote the compact Riemann surface by 
$X$ which is a special case of a $(3,3m+1)$ curve \cite{BEL2,BL1,BEL3}. The genus of $X$ is  $g=3m$.

We take $p_\infty=\infty$.
Then the affine ring $U_a$ of $X\backslash\{\infty\}$ is the space of polynomials of $x, y$ and a basis of it as a vector space is given by
\bea
&&
x^i,\hskip5mm x^i y,\hskip5mm x^i y^2,\quad i\geq 0.
\label{nonsing-basis}
\ena

We take the local coordinate $z$ around $\infty$ such that 
\bea
&&
x=z^{-3}, \quad
y=z^{-3m-1}{\hat f}(z),\quad {\hat f}(z):=\prod_{j=1}^{3m}(1-\lambda_j^3 z^3)^{1/3}.
\non
\ena

The function  ${\hat f}(z)$ is considered as a power series in $z$ by the Taylor expansion.
Expanding elements of (\ref{nonsing-basis}) in $z$ we get a frame of $\iota(U_a)$:
\bea
&&
z^{3m-3i}, \qquad, z^{-1-3i}{\hat f}(z), \qquad, z^{-3m-2-3i}{\hat f}(z)^2,\quad i\geq 0.
\label{nonsing-basis-series}
\ena

The solution corresponding to $\iota(U_a)$ can be written in terms of the multi-variate 
sigma function \cite{N2}. Let us recall it. To this end we first review the construction of the sigma 
function of $X$ following \cite{N1,N3}.

Let $f_i$, $1\leq i\leq g=3m$ be the monomials of $x,y$ defined by
\bea
&&
(f_1,...,f_g)=(1,x,...,x^m, y, xy ,...,x^{2m-1}, x^{m-1}y),
\non
\ena
and $(w_1,...,w_g)$ the gap sequence at $\infty$:
\bea
&&
(w_1,...,w_g)=(1,2,4,5,...,3m-2,3m-1,3m+2,3m+5,...,6m-1).
\non
\ena
We define $du_{w_i}$, $1\leq i\leq g$  by
\bea
&&
du_{w_i}=-\frac{f_{g+1-i}dx}{3y^2}.
\non
\ena
Then $(du_{w_i})_{i=1}^g$ becomes a basis of holomorphic one forms on $X$. The one form $du_{w_i}$ has 
a zero of order $w_i-1$ at $\infty$ and has the expansion at $\infty$ of the form
\bea
&&
du_{w_i}=\sum_{j=1}^\infty b_{ij}z^{j-1} dz,
\quad
b_{ij}=\left\{
\begin{array}{cc}
0& \text{if $j<w_i$}\\
1&\text{if $j=w_i$}
\end{array}
\right.
\label{du-expansion}
\ena
Let $\{\alpha_i\,\beta_i\}_{i=1}^g$ be a canonical homology basis, $\delta$ the Riemann's constant 
with respect to the base point $\infty$. We choose an algebraic fundamental form 
${\widehat \omega}(p_1,p_2)$ of Proposition 2 in \cite{N1}. 
We define the period matrices $\omega_i$, $i=1,2$ and $\Omega$ by
\bea
&&
2\omega_1=\left(\int_{\alpha_j}du_{w_i}\right)_{1\leq i,j\leq g},
\quad 
2\omega_2=\left(\int_{\beta_j}du_{w_i}\right)_{1\leq i,j\leq g},
\quad
\Omega=\omega_1^{-1}\omega_2.
\non
\ena
The $\alpha_j$ and $\beta_j$ integrals of ${\widehat \omega}(p_1,p_2)$ with respect to the variable $p_2$ 
are shown to be holomorphic one forms (Corollary 6 of \cite{N3}). 
With the help of this fact we define $\eta_r=(\eta_{r,ij})_{1\leq i,j\leq g}$ by
\bea
&&
\int_{\alpha_j}{\widehat \omega}(p_1,p_2)=\sum_{i=1}^gdu_{w_i}(p_1) (-2\eta_{1,ij}),
\quad
\int_{\beta_j}{\widehat \omega}(p_1,p_2)=\sum_{i=1}^gdu_{w_i}(p_1) (-2\eta_{2,ij}).
\non
\ena
The components $\eta_{r,ij}$ are the periods of certain second kind differentials.

Let $\theta[\varepsilon](z|\Omega)$ be the Riemann's theta function with the characteristics 
$\displaystyle{\varepsilon=\left[\begin{array}{c}\varepsilon'\\\varepsilon''\end{array}\right]}$,
$\varepsilon',\varepsilon''\in {\mathbb R}^g$. We consider the function
 $\theta[-\delta]((2\omega_1)^{-1}u|\Omega)$ of $u={}^t(u_{w_1},...,u_{w_g})$.

We define $\gamma\in \{w_1,...,w_g\}^m$ by
\bea
&&
\gamma=(\gamma_1,...,\gamma_m)=(6m-1, 6m-7,...,5),
\non
\ena
and set 
\bea
&&
\partial_\gamma=\partial_{\gamma_1}\cdots\partial_{\gamma_m}, 
\quad
\partial_{\gamma_i}=\frac{\partial}{\partial u_{\gamma_i}}.
\non
\ena

By Corollary 3 of \cite{N3} we have
\bea
&&
\partial_\gamma\theta[-\delta](0|\Omega):=\neq 0.
\non
\ena

\begin{defn}
The sigma function associated with $(\{du_{w_i}\}, {\widehat \omega}(p_1,p_2))$ is defined by
\bea
&&
\sigma(u)=(-1)^{\frac{m(m-1)}{2}}
\frac{\theta[-\delta]((2\omega_1)^{-1}u|\Omega)}{\partial_\gamma\theta[-\delta](0|\Omega)}
\exp\left(\frac{1}{2}{}^t u\eta_1\omega_1^{-1} u\right),
\quad
u={}^t(u_{w_1},...,u_{w_g}).
\non
\ena
\end{defn}

The sigma function has the following remarkable properties which are abesent in the Riemann's 
theta function.

\begin{theorem}\label{property-sigma}{\rm \cite{N1,N2}}
(i) The function $\sigma(u)$ does not depend on the choice of a canonical homology basis 
$\{\alpha_i,\beta_i\}$.
\vskip2mm
\noindent
(ii) The coefficients of the Taylor expansion of $\sigma(u)$ are polynomials of $\lambda_j^3$, 
$1\leq j\leq 3m$, with the coefficients in ${\mathbb Q}$.
\end{theorem}

In order to give the formula for $\tau$ function we need some more notation.
Let us write the expansion of ${\widehat \omega}(p_1,p_2)$ as 
\bea
&&
{\widehat \omega}(p_1,p_2)=
\left(
\frac{1}{(z_1-z_2)^2}+\sum_{i,j=1}^\infty {\widehat q}_{ij}z_1^{i-1}z_2^{j-1}\right)
dz_1dz_2,
\non
\ena
where $z_i=z(p_i)$. By (\ref{du-expansion}) we see that it is possible to define $\{c_j\}_{j=1}^\infty$ by
\bea
&&
\log\left(z^{-(g-1)} \sqrt{\frac{d u_{w_g}}{ d z}}\right)=\sum_{i=1}^\infty c_i\frac{z^i}{i}.
\non
\ena

By Lemma 15 of \cite{N1} we have 

\begin{prop}\label{coefficient-polynomial} 
The expansion coefiicients $ {\widehat q}_{ij}$ and $c_i$ are polynomials of $\{\lambda_j^3\}$.
\end{prop}

We set
\bea
&&
B=(b_{ij})_{1\leq i\leq g,1\leq j},
\quad
{\widehat q}(x)=\sum_{i,j=1}^\infty
{\widehat q}_{ij} x_i x_j.
\non
\ena

Then 

\begin{theorem}\label{sigma-solution}{\rm \cite{N2}}
A $\tau$ function corresponding to $\iota(U_a)$ is given by
\bea
&&
\tau(x)=\exp\left(-\sum_{i=1}^\infty c_i x_i+\frac{1}{2} {\widehat q}(x)\right) \sigma(Bx),
\non
\ena
and it is a solution of the $3$-reduced KP-hierarchy.
\end{theorem}

\section{Degeneration of $(3,3m+1)$ curves}
We consider the following  limit  of $X$:
 \bea
 &&
 \lambda_{j+m}, \lambda_{j+2m}\rightarrow \lambda_j, 
 \quad
 1\leq j\leq m.
 \label{limit-parameter}
 \ena
 The equation of $X$ tends to
\bea
&&
y^3=x\prod_{j=1}^m(x-\lambda_j^3)^3,
\hskip5mm
x=z^{-3}.
\label{trigonal-degenerate}
\ena
We assume that $\lambda_j^3$, $1\leq j\leq m$, are non-zero and mutually distinct.
It has singularities only at $Q_j=(\lambda_j^3,0)$, $1\leq j\leq m$, which is an ordinary singular point with the  multiplicity $3$ (ordinary triple point).
 In the following we denote this singular rational affine algebraic curve by $X_{\rm sing}^{\rm aff}$ and 
 by $U_a(X_{\rm sing}^{\rm aff})$ the affine ring of  $X_{\rm sing}^{\rm aff}$, that is, 
 \bea
 &&
 U_a(X_{\rm sing}^{\rm aff})={\mathbb C}[x,y]/J,
 \non
 \ena
 where $J$ is the ideal generated by $y^3-x\prod_{j=1}^m(x-\lambda_j^3)^3$.
 
 Similarly to the non-singular case  $\iota( U_a(X_{\rm sing}^{\rm aff}))$ can be defined 
 and it belongs to UGM.
 A basis of $U_a(X_{\rm sing}^{\rm aff})$ is given by (\ref{nonsing-basis}). Therefore we have 
 the corresponding basis of  $\iota( U_a(X_{\rm sing}^{\rm aff}))$ by expanding elements of  it in $z$ and 
 multiplying them by $z^{3m}$.

Let $f(z)$ be the limit of $\widehat{f}(z)$:
\bea
&&
f(z)=\prod_{j=1}^m(1-\lambda_j^3z^3).
\non
\ena
 Then the limit of the frame (\ref{nonsing-basis-series}) 
of  $\iota(U_a)$ becomes
\bea
&&
z^{3m-3i}, \hskip5mm z^{-3i-1}f(z),\hskip5mm, z^{-3m-3i-2}f(z)^2\hskip5mm i\geq 0,
\label{sing-basis-1}
\ena
which coincides with the frame of $\iota(U_a(X_{\rm sing}^{\rm aff}))$ explained above.
So (\ref{sing-basis-1}) is a frame of  $\iota(U_a(X_{\rm sing}^{\rm aff}))$. 
We show that this frame can be transformed to a fame of a generalized soliton.
To this end we need the notion of a gauge transformation. 

Let $G(z)$ be a formal power series in $z$ such that 
$G(0)=1$ and $X=(u_i(z))_{i\geq 1}$ a frame of a point of UGM. 
Then it is easy to see that $G(z)X=(G(z)u_i(z))_{i\geq 1}$ becomes a frame of a point of UGM. 
The multiplication by $G(z)$ is called a gauge transformation. Let 
\bea
&&
\log G(z)=\sum_{i=1}^\infty g_i \frac{z^i}{i}.
\non
\ena
Then
\bea
&&
\tau(x,G(z)X)={\rm e}^{\sum_{i=1}^\infty g_i x_i} \tau(x,X).
\non
\ena

Let $\omega={\rm e}^{2\pi i/3}$ and 
\bea
&&
(\lambda_1,...,\lambda_{3m})=(\lambda_1,...,\lambda_m,\omega\lambda_1,...,\omega\lambda_m,\omega^2\lambda_1,...,\omega^2\lambda_m).
\non
\ena
Notice that we redefine $\lambda_j$ ( $m+1\leq j\leq 3m$) as above. They are not relevant to 
$\lambda_j$ ($m+1\leq j\leq 3m$)  in (\ref{nonsing-3s}).

\begin{theorem}\label{main-theorem}
The following set of functions is a frame of  $f(z)^{-2}\iota(U_a(X_{\rm sing}^{\rm aff}))$. It is a frame of 
a $(6m,3m)$ generalized soliton.
\bea
&&
z^{-3m+1}\sum_{i=1}^{3m}
\left(
\frac{a_{ij}^{(0)} }{1-\lambda_i z}+\frac{b_{ij}^{(0)} }{(1-\lambda_i z)^2}
\right),
\hskip10mm
1\leq j\leq m,
\non
\\
&&
z^{-3m+1}\sum_{i=1}^{3m}\frac{a^{(r)}_{ij}}{1-\lambda_i z},
\hskip10mm
r=1,2, \hskip15mm 1\leq j\leq m,
\non
\\
&&
z^{-j}, \hskip75mm j\geq 3m.
\label{frame-deg-trigonal}
\ena
Here, for $1\leq i\leq m$,
\bea
&&
a^{(0)}_{ij}=-
\left( 
2m-j+2\sum_{l\neq i}^m\frac{\lambda_l^3}{\lambda_i^3-\lambda_l^3}
\right)
\frac{\lambda_i^{3j-8} }{3\prod_{l\neq i}(\lambda_i^3-\lambda_l^3)^2},
\non
\\
&&
a^{(0)}_{i+m,j}=\omega a^{(0)}_{ij}, \hskip5mm
a^{(0)}_{i+2m, j}=\omega^2 a^{(0)}_{ij},
\non
\ena

\bea
&&
b^{(0)}_{ij}=
\frac{\lambda_i^{3j-8} }{9\prod_{l\neq i}(\lambda_i^3-\lambda_l^3)^2},
\quad
b^{(0)}_{i+m,j}=\omega b^{(0)}_{ij}, \quad
b^{(0)}_{i+2m, j}=\omega^2 b^{(0)}_{ij},
\non
\ena

\bea
&&
a^{(r)}_{ij}=
\frac{\lambda_i^{3j-3-r} }{3\prod_{l\neq i}(\lambda_i^3-\lambda_l^3)},
\quad
a^{(r)}_{i+m,j}=\omega^{-r} a^{(r)}_{ij}, \hskip5mm
a^{(r)}_{i+2m, j}=\omega^{-2r} a^{(r)}_{ij}.
\non
\ena
\end{theorem}

In order to prove the theorem we first make a linear change of  the basis  (\ref{sing-basis-1}).

\begin{lemma}\label{lemma-sing-basis} The following is a basis of $\iota(U_a(X_{\rm sing}^{\rm aff}))$:
\bea
&&
z^{3m-3i}\quad (0\leq i\leq m-1), 
\qquad
z^{-3i}f(z) \quad (0\leq i\leq m-1), 
\non
\\
&&
z^{-3i-1}f(z) \quad (0\leq i\leq m-1), 
\qquad
z^{-3m-i} f(z)^2 \quad (i\geq 0)
\label{sing-basis-2}
\ena
\end{lemma}
\vskip2mm
\noindent
{\it Proof.} It is easy to see that the set of elements $z^{3m-3i}$, $i\geq 0$ and the set of elements 
$z^{3m-3i}$, $0\leq i\leq m-1$, $z^{-3i}f(z)$, $0\leq i\leq m-1$, $z^{-3m-3i}f(z)^2$, $i\geq 0$ are connected 
by a triangular matrix with $1$ in the diagonal entries. The set of elements $z^{-3i-1}f(z)$, $i\geq 0$ and the set of elements 
$z^{-3i-1}f(z)$, $0\leq i\leq m-1$,  $z^{-3m-3i-1} f(z)^2$, $i\geq 0$ are connected similarly.
The remaining  elements  $z^{-3m-3i-2}f(z)^2$, $i\geq 0$ of (\ref{sing-basis-1}) and  (\ref{sing-basis-2})  are the same. Thus  (\ref{sing-basis-2}) is a basis of  $\iota(U_a(X_{\rm sing}^{\rm aff}))$. $\Box$

\vskip5mm
\noindent
{\it Proof of Theorem \ref{main-theorem}}
\vskip1mm
Multiplying elements in (\ref{sing-basis-2}) by $f(z)^{-2}$ and expanding them into partial fractions 
we get (\ref{frame-deg-trigonal}). Thus it is a frame of  $f(z)^{-2}\iota(U_a(X_{\rm sing}^{\rm aff}))$. 
It follows from  
Lemma \ref{2n-k-GS} that  (\ref{frame-deg-trigonal}) is a frame of a $(6m,3m)$ generalized soliton.
$\Box$

We set
\bea
&&
b_{ij}^{(r)}=0,\hskip3mm
1\leq i\leq 3m, 1\leq j\leq m, r=1,2.
\non
\ena

The $6m\times 3m$ matrix $A$ corresponding to the frame (\ref{frame-deg-trigonal}) is given by
\bea
&&
A=\left(\begin{array}{ccc}
A^{(0)}&A^{(1)}&A^{(2)}\\
B^{(0)}&O&O\\
\end{array}
\right),
\non
\ena
where
\bea
&&
A^{(r)}=\left(a_{ij}^{(r)}+b_{ij}^{(r)}\right)_{1\leq i\leq 3m, 1\leq j\leq m},
\quad
B^{(0)}=\left(\lambda_i b^{(0)}_{ij}\right)_{1\leq i\leq 3m, 1\leq j\leq m}.
\non
\ena

The solution corresponding to the frame (\ref{frame-deg-trigonal}) of 
 $f(z)^{-2}\iota(U_a(X_{\rm sing}^{\rm aff}))$ can be computed from this matrix using the 
formula (\ref{2n-k-GSF-tau}).

\section{Solution in the case of $m=1$}
Le us consider the case of $m=1$ $(g=3)$ and write down a solution corresponding to 
 $\iota(U_a(X_{\rm sing}^{\rm aff}))$.
 
 We have 
\bea
&&
a^{(0)}_{1,1}=-\frac{\lambda_1^{-5}}{3},
\quad
a^{(1)}_{1,1}=\frac{\lambda_1^{-1}}{3},
\quad
a^{(2)}_{1,1}=\frac{\lambda_1^{-2}}{3},
\quad
b^{(0)}_{1,1}=\frac{\lambda_1^{-5}}{9}.
\non
\ena
 The matrix $A$ is a $6\times 3$ matrix given by
 \bea
 &&
 A=\left(
 \begin{array}{ccc}
 a^{(0)}_{1,1}+b^{(0)}_{1,1}&a^{(1)}_{1,1}&a^{(2)}_{1,1}\\
 \omega (a^{(0)}_{1,1}+b^{(0)}_{1,1})&\omega^2 a^{(1)}_{1,1}&\omega a^{(2)}_{1,1}\\
 \omega^2 (a^{(0)}_{1,1}+b^{(0)}_{1,1})&\omega a^{(1)}_{1,1}&\omega^2 a^{(2)}_{1,1}\\
 \lambda_1b^{(0)}_{1,1}&0&0\\
 (\lambda_1\omega)\omega b^{(0)}_{1,1}&0&0\\
 (\lambda_1\omega^2) \omega^2 b^{(0)}_{1,1}&0&0
\end{array}
\right)
=\frac{\lambda_1^{-5}}{9}
\left(
 \begin{array}{ccc}
 -2&3\lambda_1^4&3\lambda_1^3\\
 -2\omega &3\omega \lambda_1^4&3\omega^2 \lambda_1^3\\
 -2\omega^2 &3\omega^2 \lambda_1^4&3\omega \lambda_1^3\\
\lambda_1&0&0\\
 \lambda_1\omega^2 &0&0\\
 \lambda_1\omega&0&0
\end{array}
\right).
\non
\ena
For $r=0,1,2$ let 
\bea
&&
\eta_r=\sum_{j\geq 1, j\equiv r \text{ mod.$3$}}x_j\lambda_1^j,
\qquad
\eta_r'=\frac{d \eta_r}{d \lambda_1}=\sum_{j\geq 1, j\equiv r \text{ mod.$3$}}j x_j\lambda_1^{j-1},
\non
\ena
Then the $\tau$ function $\tau(x)=\tau(x, \iota(U_a(X_{\rm sing}^{\rm aff})))$ is given by
\bea
\tau(x)&=&(-1)\frac{\sqrt{3}}{27} \lambda_1^{-5}\text{e}^{-3\eta_0}
\left\{
2{\rm e}^{\frac{3}{2}(\eta_1+\eta_2)} \sin \left(\frac{\sqrt{3}}{2}(\eta_1-\eta_2)+\frac{2\pi}{3}\right)
\right.
\non
\\
&&
\left.
+2{\rm e}^{-\frac{3}{2}(\eta_1+\eta_2)}\sin \left(\frac{\sqrt{3}}{2}(\eta_1-\eta_2)-\frac{2\pi}{3}\right)
-2\sin\left(\sqrt{3}(\eta_1-\eta_2)\right)-3\sqrt{3}\lambda_1\eta_2'
\right\}.
\non
\ena
From this expression the following properties of the solution can be seen.

Firstly if  $\lambda_1$ and $x_j$, $j\geq 1$ are real then $\tau(x)$ is real. 
Secondly the variables $x_{3j}$, $j\geq 1$ only appear in ${\rm e}^{-3\eta_0}$ which means that 
 $\tau(x)$ is a solution of the 3-reduced KP-hierarchy. It is valid for any $m$
since $\tau(x)$ is a solution of the 3-reduced KP-hierarchy before taking the limit 
by Theorem \ref{sigma-solution}. Thirdly $\tau(0)=0$ which is also 
the case of any $m$.

\vskip10mm
\noindent
{\large {\bf Acknowledgments}} 
\vskip3mm
\noindent
The author would like to thank Simonetta Abenda, Takahiro Shiota and Yasuhiko Yamada for useful discussions.
He also thanks  Yuji Kodama for many valuable comments  and his interest in this work, and Kanehisa Takasaki for pointing out the similarity of the construction of generalized solitons to that of rational solutions of the KP-equation in \cite{Kr1,Kr2, Wil}.
This work is supported by JSPS Grants-in-Aid for Scientific Research No. 15K04907.

\end{document}